\documentstyle[floats,aps]{revtex}

\def\ut#1{#1\llap{\lower2ex\hbox{$\widetilde{\hphantom{#1}}$}}}

\advance\topmargin 30pt

\def\subtilde#1{#1\llap{\lower2ex\hbox{$\widetilde{\hphantom{#1}}$}}}

\begin{document} \title{Relational time in generally covariant quantum
systems: four models} 
\author{ Rodolfo Gambini$^{1}$ Rafael A.
Porto$^{1}$} 

\address{2. Instituto de F\'{\i}sica, Facultad de Ciencias,
Igu\'a 4225, esq. Mataojo, Montevideo, Uruguay.} 

\date{August 27th 2000}

\maketitle 

\begin{abstract} 
We analize the relational quantum evolution of
generally covariant systems in terms of Rovelli's evolving
constants of motion and the generalized Heisenberg picture. In
order to have a well defined evolution, and a consistent quantum
theory, evolving constants must be self-adjoint operators. We
show that this condition imposes strong restrictions to the
choices of the clock variables. We analize four cases. The first
one is non- relativistic quantum mechanics in parametrized form,
we show that, for the free particle case, the standard choice of
time is the only one leading to self-adjoint evolving constants.
Secondly, we study the relativistic case.  We show that the
resulting quantum theory is the free particle representation of
the Klein Gordon equation in which the position is a perfectly
well defined quantum observable. The admissible choices of clock
variables are the ones leading to space-like simultaneity
surfaces. In order to mimic the structure of General Relativity
we study the SL(2R) model with two Hamiltonian constraints. The
evolving constants depend in this case on three independent
variables. We show that it is possible to find clock variables
and inner products leading to a consistent quantum theory.  
Finally, we discuss the quantization of a constrained model
having a compact constraint surface. All the models considered
may be consistently quantized, although some of them do not
admit any time choice such that the equal time surfaces are
transversal to the orbits.  
\end{abstract}

\section{Introduction} An old problem in physics is how to
eliminate from our physical theories any reference to "absolute
elements". Leibnitz already used relational arguments, against
Newton's concept of absolute space and time: "...it is not
possible for two things to differ from one another in respect of
place and time alone, but it is always necessary that there
shall be some other internal difference... in general, place,
position and quantity are mere relations...". Absolute elements
are elements of the theory whose interpretation requires the
existence of things outside the dynamical system described by
the theory. The absolute acceleration in Newtonian mechanics and
the absolute interval between two events in special relativity
are absolute elements of these theories. General relativity
allows us to get rid of these elements and proposes in its place
a purely relational description of space-time.

Quantum mechanics still requires some absolute elements in order
to explain how the quantum world gives rise to events in the
(classical) measurement devices. However, it is relational in
the sense that quantum objects does not carry pre-established
answers for possible measurements, and the events are produced
during the measurements and depend upon the relation between the
quantum system and the experimental setup.

Time is also an absolute element in standard quantum mechanics.
Depending on the representation, states or operators are
parametrized by time. The choice of the time variable is
independent from the dynamics of the system, and given by the
value of some external device used as a clock. Furthermore,
there is no time operator in quantum mechanics because time has
nothing to do with the expectation value of any observable of
the quantum system. However, in the absence of a fundamental
time variable as is the case of totally covariant systems, like
general relativity, the most natural description of evolution is
relational.

Generally covariant systems are a particular kind of constrained
systems in which the motion is just the unfolding of a gauge
transformation. In other words, these theories do not have a
genuine Hamiltonian for describing the evolution of the system.
Dynamics is given by the gauge transformations generated by the
first class constraints of the theory.  In generally covariant
theories time plays the role of any other degree of freedom and
is not explicitly included within the formalism .  The canonical
variables follow orbits in phase space that may be described in
terms of arbitrary parametrizations. The arbitrary parameter
does not posses any physical significance, and therefore the
formalism is invariant under reparametrizations.

Here, we shall analize the quantum evolution of generally
covariant systems in terms of Rovelli's \cite{Ro} evolving
constants of motion and the generalized Heisenberg picture. This
procedure allows us to describe the evolution without any gauge
fixing . It is given in terms of correlations between the
original dynamical variables.  Time is identified with some
internal clock variable $T(\tau)$, and what is actually measured
is not the value of a physical variable $Q$ for certain value of
the parameter $\tau$ but the value $Q(T)$ taken by the physical
variable when the clock variable takes the value T.

Since the publication of Rovelli's seminal papers, evolving
constants have received much attention
\cite{Ku,Haj,Har,MoRoTh,Mo}.However several important problems
concerning the description of the evolution in terms of evolving
constants are still open. In the first place, contrary to what
is usually claimed, we shall see that the choice of the clock
variable is highly restricted. The fundamental limitation for
having a good clock variable arises from the nontrivial
conditions required in order to transform the evolving constants
into good, possibly unbound, self-adjoint operators of the
quantum theory. These conditions turn out to be closely related
with the transversality of the equal-time surfaces to the orbits
of the system.  Secondly, this approach was initially developed
for the study of constrained systems with one dimensional
orbits. Its extension \cite{Mo} to the general case of a system
with $N$ degrees of freedom and $n$ constraints require further
study. In fact, the $n$ independent variables required to
parametrize the evolution of the system cannot be identified as
clock variables, and further conditions on the choice of the
clock variables appear.In principle, one can choose more than
one internal degree of freedom as a clock variable.  However, we
have explored this possibility in the $SL(2,R)$ case, and we
have concluded that, one needs to use a single clock variable,
in order to have a consistent description of the evolution.  
But, we cannot rule out the possibility of having several clock
variables in more general systems. Finally, there are systems
that are not Hamiltonian, insofar as they do not admit a
decomposition of the constraint surface $\Gamma$ as a direct
product $\Gamma= \Sigma \times R$ where $R$ is the real line.
That is the case of systems with a compact constraint surface.
As it is well known, these systems are far from being well
understood.  We will show that it is possible to find time
variables leading to consistent quantizations. 

This paper is organized as follows. In Section II, we recall
what we mean by a generally covariant system, we describe two
quantization procedures for this kind of systems and we discuss
the role of the evolving constants in the description of the
time evolution. In section III, we analyze the simplest model of
a quantum generally covariant system, the parametrized
non-relativistic particle. We shall analize quantization in
terms of different choices of the clock variable, and show that
there is only one choice that leads to self-adjoint evolving
constants, the standard choice of time in non-relativistic
quantum mechanics. In section IV, we consider the case of the
relativistic particle. Quantization in terms of evolving
constants leads, for the standard choice of time, to the free
particle representation of the Klein Gordon equation. In this
representation the position is an unbound self-adjoint operator
and therefore a perfectly well defined quantum observable. We
also analize the set of admissible choices of clock variables.
In section V, we study the SL(2R) model with two Hamiltonian
constraints. This is a model with three- dimensional orbits. The
evolving constants approach is extended to this case and some
clock variables identified. In section VI we discuss the case of
systems without any natural time structure associated with
compact constraint surfaces, and show that these quantum systems
admit a consistent notion of evolution.  Section VII contains
the conclusion and some relevant considerations.

\section{Generally covariant systems}

We shall here consider finite dimensional first class totally
constrained systems. Let $\Gamma$ be the phase space of a
generally covariant system with coordinates $\{x^\mu,p_\mu\}$.
Its action S is given by

\begin{equation} 
S=\int d\tau [p_\mu {\dot x}^\mu  -  N^\alpha {\cal C}_\alpha]
\end{equation}

where ${\dot x}$ represents the derivative with respect to
$\tau$, $\mu = 1\mbox{\ldots} N$, $N^\alpha$ are Lagrange
multipliers, and ${\cal C}_\alpha$, $\alpha= 1 \ldots n$, are
the constraints. Varying $S$ with respect to $N^\alpha$ one gets
the constraint equations

\begin{equation}
{\cal C}_\alpha=0.
\end{equation}

We assume that they are first class constraints. That is, their
Poisson bracket satisfies

\begin{equation}
\{{\cal C}_\alpha, 
{\cal C}_\beta\}=f_{\alpha\beta\gamma}{\cal C}_\gamma
\end{equation}

where $ f_{\alpha\beta\gamma}(x,p)$ are the structure functions.
In most of the models we shall consider structure constants.

A basic notion of our description will be that of observable. An
observable is a function, in general complex, in phase space
such that its Poisson Bracket with the constraints vanishes. The
observables are the gauge invariant quantities of the system. In
a generally covariant system, the observables are constants
along the orbits. That is the reason why, in this case, they are
also called perennials.  The perennials form a Poisson algebra
with involution. The evolving constants are a particular kind of
observables\cite{Ro} defined as follows. Let us assume for
simplicity that the constrained system is defined by a single
constraint ${\cal C}$ on a phase space $x^\mu,p_\mu$, $\mu= 1
\ldots N$ and let us take one of the variables, say $x^1$ as a
clock variable. Then we want to introduce observable quantities
that represent the evolution of $x^i$ $i= 2 \ldots N$ and
$p_\mu$ in terms of $x^1$.  To do that, we consider a
one-parameter family of observables. Each observable of the
family represents the value of $x^i$ at a different value, say
$T$, of the clock variable $x^1$. Let us call these observables
$X^i(T)$; they satisfy

\begin{eqnarray}
\{X^i(T,x^\mu,p_\mu),{\cal C}(x^\mu,p_\mu)\} &=& 0 \label{ham}\\  
X^i(T, x^1=T, x^2, x^3 \ldots p_N) &=& x^i.  \label{init}
\end{eqnarray}

The first equation imposses that $X^i$ is constant along the
orbits, the second determines the value taken by the observable
along the orbit.
 
For instance, take the presymplectic system associated with the
non relativistic free particle. The phase space is four
dimensional with coordinates $(x^0,x^1, p_0,p_1)$, and the
physical trajectories are contained in the constraint surface
${\cal C}=p_0+p^2/{2m}$. The evolving constant that describes
the evolution of $x^1$ in terms of $x^0$ is given by

\begin{equation}
X^1= x^1- {p \over m} (x^0-T)
\end{equation}   

It is immediate to check that equations (\ref{ham}) and
(\ref{init}) are satisfied. Notice that the evolving constant
may be interpreted as the expression of a dynamical variable in
terms of the initial conditions, which are perennials, and a
value of the clock variable. When the clock variable evolves, it
reproduces the classical evolution of the dynamical variables.

As we shall see, generally covariant systems with more than one
constraint require more than one independent variable to
coordinate the orbits. A subset of these variables may be taken
as clock variables while the others may be interpreted as
spatial coordinates.

In what follows we shall assume that we have identified a
complete set of perennials such that any classical perennial is
a function of the elements of this set. Let us consider the
Poisson algebra with involution $A$ generated by this set.
Evolving constants are generically functions of the complete set
of perennials, but they do not necessarily belong to their
algebra. The starting point of our description of the time
evolution of a generally covariant system will be the extended
Poisson algebra with involution ${B^{(*)}} _{phys}$ that
contains $A$ and a suitable set of evolving constants. Within
this approach, the resulting evolution will depend on this
choice of evolving constants.

There are several alternative procedures of quantization of
totally constrained systems. The resulting quantum theory may
differ depending on the procedure.  These methods are not always
equivalent due to factor-ordering problems or nontrivial global
issues.  Here, we are going to consider, two types of
quantization procedures. The first type of procedure consists in
quantizing first the kinematical variables and after that
imposing the constraints. Algebraic Quantization\cite{Ash,AshTa}
and Refined Algebraic Quantization\cite{AshLeMaMoTh,GiMa} are
procedures of this type. In the second type of procedure, the
reduced phase space quantization, one is only concerned with the
observable quantities. Thus one starts from the classical
Poisson algebra with involution ${B^{(*)}}_{phys}$, and look for
a quantum representation of the corresponding operators
belonging to ${{\cal B}^{(*)}}_{phys}$ on a Hilbert space ${\cal
H}'_{phys}$ such that their reality properties be preserved by
the quantization Depending on the model,one method could be more
convenient than the other.
 
As it is technically more involved, we will start by making a
brief introduction to the refined algebraic quantization
procedure. We refer the interested reader to the original
references \cite{AshLeMaMoTh,GiMa}.  One starts by introducing
operators $\hat x$ and $\hat p$ on an auxiliary Hilbert space
${\cal H}_{aux}=L^2(R^N)$. Let ${\hat x}^\mu$ be multiplicative
and ${\hat p}_\mu$ act by ($-i$ times)  differentiation, so that
all the operators are self-adjoint on ${\cal H}_{aux}$.  One
represents the constraints ${\cal C}_\alpha$ as self-adjoint
operators ${\hat {\cal C}}_\alpha$, on ${\cal H}_{aux}$. Now,
consider the quantum version ${{\cal B}^{(*)}}_{phys}$ of the
classical algebra ${B^(*)}_{phys}$, whose elements are operators
(perennials) on ${\cal H}_{aux}$ which commute with the
constraints ${\hat {\cal C}}_\alpha$.  They should be linear,
but not necessarily bounded operators. From a physical point of
view, the identification of a suitable family of perennials is a
key step. In particular, as it was already mentioned, it
involves the choice of a set of evolving constants in order to
describe the evolution of the system.

The solutions of the constraints define a linear space $L$. As
generically the constraints will not be bounded, it will be
imposible to identify $L$ with ${\cal H}_{aux}$. For example, if
the spectrum of any given constraint is purely continous, then
it has no normalizable eigenvectors on ${\cal H}_{aux}$, and
therefore the elements of $L$ will not be contained in ${\cal
H}_{aux}$. In fact, the elements of $L$ will be distributions
belonging to the dual of some dense subspace $S_L$ of ${\cal
H}_{aux}$. One imposes the following conditions on $S_L$
\cite{GiMa}: i)$S_L$ is contained in the domain of each
constraint ${\hat {\cal C}}_\alpha$.  ii) $S_L$ is left
invariant under the action of each constraint.  These two
conditions ensure that the solutions of the constraints belong
to the dual of $S_L$. Furthermore one requires that the elements
$\hat A$ of ${{\cal B}^{(*)}}_{phys}$ be such that: iii)the
domain of $\hat A$ and ${\hat A}^\dagger$ contains $S_L$, and
iv) $\hat A$ and ${\hat A}^\dagger$ leave $S_L$ invariant. These
last two conditions imply that it is possible to induce an inner
product in the physical space such that the reality properties
of the physical operators of ${{\cal B}^{(*)}}_{phys}$ are
ensured.

The physical Hilbert space is introduced as follows: One defines
an inner product in some subspace $M$ of $L$ in such a way that
its completion is the physical Hilbert space ${\cal H}_{phys}$.
It is always possible under the aforementioned hypothesis to
induce an inner product in $M$ such that the operators of
${{\cal B}^{(*)}}_{phys}$ corresponding to real classical
perennials are self adjoint. The explicit construction of the
inner product is given in Refs \cite{AshLeMaMoTh,GiMa}

As we shall see in what follows, evolving constants are a
powerful tool to study the evolution of generally covariant
systems. In particular, they allow us to compare the evolution
resulting from the use of different clock variables simply by
including in ${{\cal B}^{(*)}}_{phys}$ the evolving constants
corresponding to the different choices.

The main assumption is that the operators belonging to ${{\cal
B}^{(*)}}_{phys}$ corresponding to real quantities are
self-adjoint. Any choice of clock variable leading to self
adjoint evolving constants is valid. We shall show that,
contrary to what is usually claimed, this condition imposses
strong restrictions to the choices of clock variables. In fact,
although general procedures for systematically ordering
observables exist \cite{Moy,Ko} and they ensure that the
resulting operators are symmetric, in the case of unbound
operators this is not enough to ensure that they are also
self-adjoint \cite{AkGl}. The choice of the clock variables not
only fixes the rate of the running clocks, but also determines
the set of simultaneous events.  An "equal-time" surface
$\Sigma$ is called transversal to the orbits, if each orbit
intersects $\Sigma$ in one and only one point and the tangent
vectors to the orbit are independent from the tangent vectors to
$\Sigma$. We shall see that transversality is not required to
have self-adjoint evolving constants, and a weaker condition is
sufficient to ensure a consistent description of the quantum
evolution.

Notice that the quantum description resulting from this
procedure corresponds to a Heisenberg representation, i.e., the
observables evolve in time. However, one does not assume the
existence of a Hamiltonian such that for any perennial $Q(T)$
the following equation holds

\begin{equation}
ih {d Q(T)\over  dT} = [Q(T),H]     
\end{equation} 

The resulting quantum description in terms of evolving constants
is relational in the sense that it assigns probabilities to
events produced in measurement devices when certain dynamical
variables taken as clocks take a given value.  Depending on the
choice of clock variable and on the physical quantity under
consideration (represented by an evolving constant) we will have
a different experimental setup and a different procedure for the
synchronization of the clocks. One can describe the same
experimental setup in different states of motion, by changing
the parameterization of the evolving constants. It is even
possible, by changing the clock variable, to introduce two
different orders for the same set of events. If both orders lead
to well defined quantum theories, their predictions should be
consistent.

\section{Parametrized non-relativistic quantum mechanics}

Parametrized non-relativistic quantum mechanics admits a natural
time structure. As it was already shown by Rovelli
\cite{Ro42,Ro} , the standard choice of the clock variable
$T=x^0$, leads to predictions that agree with usual
non-relativistic quantum mechanics. However, the method of
evolving constants admits, in principle, a much wider variety of
choices for the clock variable. In principle, these choices are
only restricted by the requirement that the corresponding
evolving constants are well defined self-adjoint operators on
some Hilbert space allowing to preserve the probabilistic
structure of the resulting quantum theory. In Rovelli's words,
the idea is that "the time axiom can be dropped without
compromising the other axioms of the probabilistic
interpretation of the theory." Here we intend to analyze whether
other choices of the clock variable meet these requirements. In
what follows we will show that there is strong evidence
suggesting that the only allowed time choice of the form
$T(x^i)$ is the standard one. \footnote{this problem was also
analized by Hartle \cite{Har} who also concluded based on
physical arguments that the allowed time functions should be
restricted for the predictions to coincide with those of the
usual quantum theory.}
 
Let us consider the elementary case of a non-relativistic free
particle in one dimension. Let $\Gamma$ be the phase space of
the system with coordinates $(x^0,x)$ and momenta $(p_0,p)$. Its
action is given by

\begin{equation} 
S=\int d\tau [p_0 {\dot x}^0+ p{\dot x}  -  
N(p_0+ \textstyle{{p^2}\over{2m}}) ]
\end{equation}
that leads to the Hamiltonian constraint

\begin{equation} 
{\cal C} \equiv p_0+
\textstyle{{p^2}\over{2m}}=0 
\end{equation} 

which defines the
constraint surface $\Gamma'$.  A complete set of classical
perennials is given by $p$ and $x-x^0p/m$. Now we choose the
auxiliary Hilbert space to be ${\cal H}_{aux}= L^2(\mbox{\rm
R}^2,dp_0dp)$ and represent $x^\mu$ and $p_\mu$ by the usual
expressions of the momentum representation, so that the
corresponding operators are self-adjoint.  Let us now choose
clock variables and the corresponding evolving constants that
will be used to describe the evolution, and include them among
the elements of ${{\cal B}^{(*)}}_{phys}$.
	
Our first choice will be $T=x^0$, then ${{\cal B}^{(*)}}_{phys}$
will include the perennials $\{\hat q := \hat {x-x^0p/m}, \hat
p, \hat X(T):= \hat q + {\hat p T} /m \}$ The spectrum of $\hat
X(T)$ will characterize the possible outcomes of a measurement
of $\hat X$ when $x^0$ takes the value $T$. Our physical states
will be associated with generalized eigenstates of ${\cal C}$
with vanishing eigenvalue. These states may be considered as
distributions acting on a dense subspace $S_L \subset {\cal
H}_{aux}$ In this case, the natural choice for $S_L$ is the
space of smooth functions with compact support. Notice that, for
this choice of the clock variable, $S_L$ satisfies the
requirements established in the previous section.  The physical
states are

\begin{equation}
\psi(p,p_0)= \delta(p_0 + \textstyle{{p^2}\over{2m}})f(p)
\end{equation}

The perennials $\hat p, \hat q, \hat X(T)$ are self-adjoint on
${\cal H}_{aux}$ and satisfy the following generalized
eigenvalue equations:

\begin{eqnarray}
\hat p \psi_{p_1}(p,p_0)&=& p \psi_{p_1}(p,p_0)= p_1\psi_{p_1}(p,p_0)\\
\hat q \psi_{q_1}(p,p_0)&=& (i \textstyle{\partial \over {\partial p}}-ip/m 
\textstyle{\partial \over {\partial{p_0}}})\psi_{q_1}(p,p_0) = 
q_1\psi_{q_1}(p,p_0)\\
\hat X(T) \psi_{x_1,T}(p,p_0)&=& 
[i \textstyle{\partial\over{\partial p}}-ip/m
\textstyle{\partial \over {\partial{p_0}}}+ {T p }/m] \psi_{x_1,T}(p,p_0)
= x_1\psi_{x_1,T}(p,p_0),
\end{eqnarray}

The solutions belonging to the kernel of the constraint are

\begin{eqnarray}
\psi_{p_1}(p,p_0)&=& \delta(p_0+\textstyle{p^2\over 2m})\delta(p-p_1)\\
\psi_{q_1}(p,p_0)&=& \delta(p_0+\textstyle{p^2\over 2m})\exp{ipq_1}
\end{eqnarray} 

and

\begin{equation}
\psi_{x_1,T}(p,p_0) = \delta(p_0+{\textstyle{p^2\over 2m}})
\exp{ i(px_1 - {\textstyle{{p^2}\over{2m}}}T)}. \label{eigenf}
\end{equation}

It is now possible to introduce \cite{AshLeMaMoTh} an inner
product in the physical space of solutions of the constraint.
One considers solutions of the form $\psi(p,p_0)=
\delta(p_0+p^2/2m)f(p,p_0)$, with $f(p,p_0) \in S_L$. Then, the
inner product is given by:

\begin{equation}
<\psi_1|\psi_2>_{phys} = 
\int_{-\infty}^\infty dp\int_{-\infty}^\infty 
dp_0   \delta(p_0+\textstyle{p^2\over 2m}){f^*}_1(p,p_0)f_2(p,p_0)= 
\int_{-\infty}^\infty dp {{\tilde f}^*}_1(p){\tilde f}_2(p) 
\label{inner}
\end{equation}

where ${\tilde f}(p)=f(p,-p^2/2m)$. Then, by considering the
completion of $S_L$ one gets a Hilbert space ${\cal H}_{phys}$
with the standard inner product in the momentum representation.
All the elements of ${{\cal B}^{(*)}}_{phys}$ corresponding to
real quantities will be trivially defined self-adjoint operators
in this space. They will be given by ${\hat q}\psi= i {\partial
\psi}/\partial p, {\hat p}\psi = p\psi ,\;\mbox{\rm and}\; X(T)=
\hat q + {p/m}T$.  Notice that in this simple case, if we had
followed the second procedure and taken as starting points the
algebra of observables we would have recovered in a simpler way
the same quantum operators and the inner product (\ref{inner}).

Summarizing, the choice of clock variable $T=x^0$ leads to the
standard form of the quantum free particle in the Heisenberg
representation. In particular one may recover from
Eqs(\ref{eigenf}) and (\ref{inner}) the standard transition
amplitude

\begin{equation}
<\psi_{x,T}|\psi_{x',T'}>=  [{2 \pi i (T-T')\over m }]^{-1/2}
\exp{{i m(x-x')^2 \over 2 (T-T')}}.\label{prop}
\end{equation}

Now, we would like to analyze other choices of the clock
variable for this simple model and determine whether they lead
to self-adjoint evolving constants and admit a probabilistic
interpretation. Let us first consider the choice of the clock
variable $x = T$. The corresponding evolving constant is
$X^0(T)\equiv ({{-m}/p})(q-T)$. \footnote{This choice has been
extensively analyzed in Ref.\cite{GrRoTa}} It obviously commutes
with the constraint ${\cal C}$ and $X^0(x=T)= x^0$. The
inclusion of ${\hat X}^0(T)$ in ${{\cal B}^{(*)}}_{phys}$ will
immediately lead to difficulties. The space $S_L$ of smooth
functions of constant support does not satisfy the conditions
iii) and iv) of the previous section. Due to the division by
$p$, $S_L$ is not contained in the domain of $\hat X^0(T)$. Only
the wave functions with zero amplitude for $p=0$ might belong to
${\cal H}_{aux}$.  The resulting operator is not self-adjoint
and does not admit self-adjoint extensions. Let us show that
explicitly by following the reduced phase space quantization
procedure. The standard inner product in momentum representation
ensures that the complete set of perennials $\hat q$, and $\hat
p$ is self-adjoint. Let us consider the symmetric form of the
evolving constant ${\hat X}^0 (T)$ in the momentum
representation given by:

\begin{equation} 
\hat X^0(T)\psi(p) = [{{-1}\over 2} ({m \over p} {i \partial
\over \partial p}+  {i \partial \over \partial p} {m \over p}) +
{{mT}\over p}]\psi(p)  
\end{equation}

In order to find out whether this operator is self-adjoint or at
least has self-adjoint extensions, we are going to use the
following simple and powerful method \cite{Si,Wi}: Check the
dimensionality of the two subspaces: ${\cal K}_-= \hbox{\rm
ker}(X^0(T)-i)$ and ${\cal K}_+= \hbox{\rm ker}(X^0(T)+i)$. If
they do not have the same dimensionality, the operator is not
self-adjoint and has no self-adjoint extensions.

It is immediate to see that

\begin{equation}
\hbox{\rm dim}\; \hbox{\rm ker}(X^0(T)+i)= \hbox{\rm dim}\{c 
\sqrt{{p}\over{2 \pi}}
\exp(-ipT -{p^2\over 2m})\}=1,
\end{equation}

because the elements of this kernel belong to
$L^2[-\infty,\infty,dp]$, while

\begin{equation}
\hbox{\rm dim} \hbox{\rm ker}(X^0(T)-i)= \hbox{\rm dim}\{c 
\sqrt{{p}\over{2 \pi}}
\exp(-ipT +{p^2\over 2m})\}=0.
\end{equation}

Thus, the evolving constant $X^0$ is not self-adjoint, and it
does not correspond to any observable quantity of the quantum
theory. In physical terms it is not possible to assign
probability amplitudes to the generalized eigenvalues of the
operator. The physical origin of this problem is the following:
The equal "time" surfaces $x = T$ are not transversal to the
orbits and the measurement devices lying at different values of
$x^0$ are causally connected. In particular, a particle may be
found, at equal $T$, at two different values of $x^0$ as it can
be seen by computing the inner product of two eigenfunctions
corresponding to different values of $x^0$. In fact, the
eigenfunctions of $X^0(T)$ with eigenvalue $x^0$ are given by:

\begin{equation}
\psi_{T, x^0}(p) = \sqrt{{p}\over{2 \pi}}
\exp(-ipT + i{p^2\over 2m}x^0)
\end{equation} 

Therefore the functions corresponding to different
eigenvalues are not orthogonal. In fact

\begin{equation}
<\psi_{T,x^0}|\psi_{T,{x^O}'}> = 
{1\over 2 \pi}\int_{-\infty}^\infty 
dp |p| \exp{i \textstyle{p^2\over 2m}({x^0}'-x^0)}
\end{equation}

and introducing the variable $p_0={p^2\over 2m}$ one gets

\begin{equation}
<\psi_{T,x^0}|\psi_{T,{x^0}'}> = {m \over 2\pi}\int_{0}^\infty dp_0 
\exp{ip_0({x^0}'-x^0)}=
{m \over 2}[\delta({x^0}'-x^0) -2 \pi i {\cal P}{1\over {{x^0}'-
x^0}}], \label{prod}
\end{equation}
 
where ${\cal P}$ is the principal part. One may wonder whether
requiring the evolving constant to be self-adjoint is too
stringent when we are considering operators associated to
causally connected measurements. After all, as we are taking
$x=T$ the states are prepared by measurements performed for all
the possible values of $x^0$ at a given $x$, and the standard
free particle propagator (\ref{prop})  leads to nonvanishing
transition amplitudes between points with coordinates $x,x^0$
and $x,{x'}^0$. However, we don't see any consistent procedure
for the assignment of probabilities when the operators are not
self-adjoint. In fact, the natural way for the assignment of
probabilities to the measurement of a quantum observable, is via
the quadratic form $<\phi|A|\phi>$. This expression represents
the mean value of the Heisenberg observable only if $A$ is a
self- adjoint operator with a spectral decomposition for any
value of the time variable $T$. If that is not the case, one
looses the natural interpretation for the proper values as a
possible outcome of the measurement. Notice, furthermore, that
if one simply attempts to identify the inner product
(\ref{prod}) with the free particle kernel (\ref{prop})  
evaluated for $x=x'=T$, one notices that the inner product does
not reproduce the standard transition amplitude.

Other choices of the clock variable leading to equal time
surfaces which are not transversal to the orbits also lead to
the same kind of difficulties. For instance the choice
$T=x^0-ax$, where $a$ is an arbitrary constant, corresponds to
the evolving constant

\begin{equation}
X(T) = {{q+\textstyle{p \over m}T} \over {1- a\textstyle{p\over m}}}
\end{equation}    

and leads to a non self-adjoint operator. Again the smooth
functions of $p$ with compact support lay outside the domain of
$\hat X$ and the kernels of $\hat X \pm i$ have different
dimension. Notice that this choice of clock variable do not
differ from the standard choice in the rate of the running
clocks, but define different equal time surfaces. In fact this
choice corresponds to an array of measurement devices at rest,
where the clocks have been synchronized on the surface $x^0=ax$,
and therefore "simultaneous"  measurements of the position are
causally connected by orbits with speed $1/a$.

One could have considered an array of measurement devices in
uniform motion. In order to do that, one may introduce the
Galileo transformation $x'= x+ vx^0$ and ${x'}^0= x^0$. Then if
one defines an evolving constant $\hat X'$ associated with $x'$
and choose the clock variable $x^0=T$, one gets a perfectly well
behaved self-adjoint operator and a consistent quantum
mechanical description. It is also possible to extend this
analysis to the case of non-relativistic particles under the
action of simple conservative forces. One is led to the same
conclusion for all the cases considered: The requirements for
the existence of a consistent quantum mechanical description of
the non relativistic particle are fulfilled if, and only if, the
choice of the clock variable leads to equal time surfaces which
are transversal to the orbits.  As we shall see in the following
sections, one should not hasten to conclude, from these simple
models with a natural time structure, that in the case of
generally constrained systems without a natural time structure
of transversal surfaces it will be impossible to have a
consistent quantum mechanical description of the evolution.

\section{Parametrized relativistic quantum mechanics}

As a second example of systematic application of the evolving
constants technique, we are going to study the quantization of
the relativistic particle. This system is conceptually more
involved than the previous one due to several issues. In the
first place, while the non-relativistic particle only admits a
natural time structure, the relativistic system should admit a
wider class of clock variables associated to space like
slicings. Secondly, the phase space of the relativistic particle
with quadratic Hamiltonian constraint has two disconnected
sectors. When one attempts to quantize the theory including both
sectors, for instance using the Klein Gordon procedure, one runs
into difficulties. The inner product is not positive definite,
the position is not self-adjoint, and when the particle is
subject to external forces which are not slowly varying on a few
Compton wavelengths, the system does not admit any probabilistic
interpretation. We are going to show that the evolving constants
technique sheds new light on these phenomena.

Let us briefly consider, in the first place, the simplest case
of a relativistic particle with positive energy. The action is
given by

\begin{equation} 
S=\int d\tau [p_0 {\dot x}^0+ p{\dot x}  -  N(p_0+ \sqrt{p^2+m^2}) ]
\end{equation}

We follow steps parallel to the case of the non-relativistic
particle. Again the auxiliary Hilbert space will be ${\cal
H}_{aux}= L^2(\mbox{\rm R}^2,dp_0dp)$ A complete set of
classical perennials is given by $p$ and
$q:=x-x^0p/{\sqrt{p^2+m^2}}$. In this case, the system admits
several consistent choices of clock variables. The first and
trivial choice is $x^0=T$. We take as an evolving constant the
one giving the value of the coordinate $x$ when the clock takes
the value $T$. It is given by $X(T)= q+ {pT}/\sqrt{p^2+m^2}$,
which should be included together with the perennials $q$ and
$p$ in $B^{(*)}_{phys}$. These quantities are trivially
quantized on ${\cal H}_{aux}$. They satisfy conditions i) to iv)
when acting on the space of smooth functions with compact
support $S_L$. It is therefore not surprising that they be
self-adjoint operators. For instance, it may be immediately
checked that the dimensionality of the subspaces ${\cal K}_+$
and ${\cal K}_-$ is the same and equal to zero. This is
therefore a consistent choice of time, and reproduces the
standard results for a relativistic scalar particle with
positive energy. One can, for instance, compute the generalized
eigenfunctions of $\hat X(T)$, which satisfy orthonormality and
closure relations and lead to the standard positive energy
propagator of the relativistic particle.

This model admits other consistent choices of clock variable or
evolving constants. For instance, from the choice of clock
variable $T=x^0-ax$, where $a$ is a real constant, one may
define two evolving constants

\begin{equation}
X(T)= 
{q+ { pT \over \sqrt{p^2+m^2}}\over{1-{{ap}\over \sqrt{p^2+m^2}}}}
\end{equation}
 
and

\begin{equation}
X^0(T)= {{-a q+ T}\over{1+{{ap}\over \sqrt{p^2+m^2}}}}.
\end{equation}

The first one takes the value $x$ for $T=x^0+ax$, while the
second one takes, for the same value of the clock variable, the
value $x^0$. It is not difficult to check that at the quantum
mechanical level these choices lead to consistent theories for
any value of $a \leq 1$. We will give here the explicit proof of
the self-adjointness of the $\hat X(T)$ operator. The reader may
reproduce the analysis for $X^0(T)$. By using the reduced phase
space quantization, one immediately gets the symmetric operator
corresponding to the evolving constant $\hat X(T)$. It is

\begin{equation}
{\hat X}(T)\psi(p) = [{i\over 2}{\partial \over \partial p}{1\over
{1-{pa\over \sqrt{p^2+m^2}}}}+
 {i\over 2}{1\over
{1-{pa\over \sqrt{p^2+m^2}}}}{\partial \over \partial p} 
+{pT\over{\sqrt{p^2+m^2} -pa}}]\psi(p).
\end{equation}

Looking at the dimensions of the two subspaces ${\cal K}_+$ and
${\cal K}_-$ one gets:

\begin{equation}
\hbox{\rm dim}\; \hbox{\rm ker}(\hat X(T)+i)= \hbox{\rm dim}
\{c \sqrt{1-{pa\over \sqrt{p^2+m^2}}} 
\exp(i \sqrt{p^2+m^2} T +[a\sqrt{p^2+m^2}-p])\}=0
\end{equation}

and 

\begin{equation}
\hbox{\rm dim}\; \hbox{\rm ker}(\hat X(T)-i)= \hbox{\rm dim}
\{c \sqrt{1-{pa\over \sqrt{p^2+m^2}}} 
\exp(i \sqrt{p^2+m^2} T -[a\sqrt{p^2+m^2}-p])\}=0
\end{equation}

for $a \leq 1$, because these elements do not belong to
$L^2[-\infty,\infty,dp]$. For $a > 1$ the dimensions are
different, while the kernel ${\cal K}_+$ has dimension zero, the
second kernel ${\cal K}_-$ has dimension 1.

The self-adjoint evolving constants ${\hat X}(T)$ and ${\hat
X}^0(T)$ are associated with the production of events in the
same measurement devices. In both cases, we choose to perform a
T simultaneous measurement that means that we are opening the
detectors synchronised on the T-surface. However, while in the
first case the values of the position of the event are
registered, in the second what is registered is the proper time
of the corresponding measurement devices.

In both cases, the simultaneity surfaces $x^0+ax=T_0$ coincide,
and for $a \leq 1$ they are transversal to the orbits. Thus, as
in the non-relativistic case, in order to have a consistent
quantum mechanical description one need to choose clock
variables leading to equal time surfaces transversal to the
orbits. This seems to be a general feature of the evolving
constant technique when applied to systems having a natural
Hamiltonian evolution.

When both sectors of the relativistic free particle phase space
are included at the same time, the system does not have a
natural Hamiltonian evolution. As it is well known, the standard
Klein Gordon quantization is far from satisfactory, leading to
an inner product which is not positive definite. Generically, it
is not consistent with a probabilistic interpretation. Here, we
would like to analyze whether the description of the evolution
in terms of the evolving constants which in principle does not
require any Hamiltonian structure could give a new insight to
this old problem.

Let $\Gamma$ be the phase space of the system with
coordinates $(x^0,x)$ and momenta $(p_0,p)$. Its action is given by

\begin{equation} 
S=\int d\tau [p_0 {\dot x}^0+ p{\dot x}  -  N({p_0}^2 - p^2 -m^2) ]
\end{equation}

that leads to the Hamiltonian constraint

\begin{equation} 
{\cal C} := {p_0}^2 - p^2 -m^2=0
\end{equation}

which defines the constraint surface $\Gamma'$.  A complete set
of classical perennials is given by $p$, $q:=x+x^0p/{p_0}$ and
$\epsilon := {p_0}/\sqrt{p^2+m^2}=\mbox{\rm sg}(p_0)$. Now we
choose the auxiliary Hilbert space to be ${\cal H}_{aux}=
L^2(\mbox{\rm R}^2,dp_0dp)$ and represent $x^\mu$ and $p_\mu$ by
the usual expressions of the momentum representation, so that
the corresponding operators are self-adjoint.  Let us now choose
clock variables and the corresponding evolving constants that
will be used to describe the evolution, and include them among
the elements of ${{\cal B}^{(*)}}_{phys}$.

We will analyse the usual choice $T=x^0$; then ${{\cal
B}^{(*)}}_{phys}$ will include the operators $\{\hat q, \hat p,
\hat \epsilon,\; \hbox{\rm and}\; \hat X(T):= \hat q + {\hat p
T} /{p_0} \}$ The spectrum of $\hat X(T)$ will characterise the
possible outcomes of a measurement of $\hat X$ when $x^0$ takes
the value $T$. It is given by the eigenvalue equation:

\begin{equation}
{\hat X}(T)\psi_{x_1,T}(p,p_0)= 
[i  {\partial \over \partial p} + i{p\over
{p_0}}{\partial \over \partial p_0} - 
i {p \over 2{p_0}^2} +{p T\over p_0}]
\psi_{x_1,T}(p,p_0)=x_1\psi_{x_1,T}(p,p_0) 
\end{equation}

Notice that, as $[\hat \epsilon,\hat X(T)]= [\hat \epsilon, \hat
p] =0$, one may consider simultaneous eigenfunctions of $\hat X$
and $\hat \epsilon$ The generalized eigenfunctions belonging to
the kernel of the constraint equation are,

\begin{equation}
\psi_{x_1,T,\epsilon_1}(p,p_0)=
 \delta_{\epsilon_1 \epsilon}\delta({p_0}^2-p^2-
m^2)\sqrt|p_0|\exp{(-ipx-ip_0T)}
\end{equation}

where $\epsilon={{p_0}/\sqrt{p^2+m^2}}$.  As in the
non-relativistic case, the inner product in ${\cal H}_{aux}$
induces\cite{AshLeMaMoTh} an inner product in the physical state
space.

\begin{equation}
<\psi|\psi'>_{phys}=\int dp dp_0 \delta({p_0}^2-p^2-
m^2)\psi^*(p,p_0)\psi'(p,p_0)\label{inuno}
\end{equation}

With this inner product, the operators $\hat X(T)$ and $\hat
\epsilon$ are self-adjoint and the corresponding eigenfunctions
orthonormal. Notice, however, that with this inner product the
expectation value of the energy $<\psi|-p_0|\psi>$ may take
negative values. If one is interested in avoiding states with
negative energy, one may introduce a modified inner product
given by:

\begin{equation}
<\psi|\psi'{>^M}_{phys}=-\int dp dp_0 \delta({p_0}^2-p^2-m^2)\hbox{\rm 
{Sg}}(p_0)\psi^*(p,p_0)\psi'(p,p_0)\label{inmod}
\end{equation}
 
where $-\hbox{\rm {Sg}}(p_0)$ is the sign of the energy. Thanks
to the commutation of $\hat X(T)$ with $\hat \epsilon$, the
self-adjointness of the evolving constant $\hat X$ is preserved,
but the new inner product is not positive definite. In fact the
resulting formulation coincides with the free particle
representation \cite{FeVi} of the Klein Gordon theory. The
quantum version of the evolving constant turns out to be
automatically self-adjoint and coincides with the Newton-Wigner
\cite{Wig} position operator.
 
From the relational point of view adopted in this paper, each
system is treated as a closed system. It that sense, the
relativistic particle is considered as a cosmological system
\footnote{A similar cosmological interpretation of the
relativistic particle was proposed by Marolf\cite{Ma}}, and
makes no reference to an order in time, there is no before or
after, only events labelled by different values of the
parameter. Within this context, the first inner product given by
(\ref{inuno}) is more natural. It leads to a consistent
quantization with a positive definite inner product and self
adjoint observable quantities, treats on an equal footing the
particles with positive energy, travelling forwards in time, and
the particles with negative energy, travelling backwards.
 
If one wants to include interactions one needs to start from the
evolving constants of the system including the interaction. It
is well known that the relativistic systems present a highly
anomalous behaviour, the so-called Klein paradox, when the
system is coupled with an external field with abrupt variations
at distances of the order of the Compton wavelength. Let us
conclude this section with a brief discussion of the evolving
constant formalism applied to a particle interacting with an
step-function potential. The Hamiltonian constraint is given by

\begin{equation}
{\cal C} = (p_0-\phi(x))^2 - p^2 -m^2
\end{equation}

with $\phi(x)= -c U(x-a)$, $U$ is the step function and we will
take $c > m$. In this case one can see that the evolving
constant $\hat X(T)$ does not commute with the sign of the
energy characterized by $\hat \epsilon$. Thus, even though $\hat
X$ is self-adjoint with the symmetric inner product that is the
extension of (\ref{inuno}) to the case of the relativistic
particle in an external field, it will be impossible to define a
self-adjoint operator with the modified inner product
(\ref{inmod})  that corresponds, in the free particle case, to
the Klein Gordon representation. Thus, from the viewpoint of the
evolving constants formalism, the Klein paradox can be
associated to the impossibility of defining a self-adjoint
operator for the position when step potentials are present. To
conclude the analysis of the relativistic particle, the evolving
constant procedure allows treating both sectors of the classical
phase space at the same time. From a purely relational and
therefore cosmological point of view, the theory with the first
inner product should be preferred because it admits a
probabilistic interpretation, even when the system interacts
with external fields. This theory gives a symmetric treatment to
states with positive and negative "energy" and should be
reinterpreted as the cosmological description of a "Universe"
classically described by the relativistic particle. Kuchar
\cite{Kuca} have shown that the relativistic particle moving in
a Riemannian spacetime presents a multiple choice problem: There
are different admissible time functions leading to incompatible
quantum mechanical schemes. We are now studying these systems
making use of the point of view advocated in this paper.
 
\section{The SL(2,R) model with two Hamiltonians}

Up to now, we have analyzed the relational formalism in systems
with a single Hamiltonian constraint and one-dimensional orbits.
The SL(2,R) model \cite{MoRoTh} is a constrained system with two
Hamiltonian constraints that have non-vanishing Poisson bracket
with each other. The model has three constraints and their
algebra mimics the structure of the constraint algebra of
general relativity. Although a complete set of evolving
constants associated with the dynamical variables of the systems
has been identified \cite{Mo}, it has not been possible, up to
now, to identify clock variables that allow to promote this
quantities to well defined self-adjoint operators. We shall see
that algebraic quantization and reduced phase space quantization
lead, in this case, to different quantum theories. The second
procedure will allow us to introduce a consistent time structure
and to promote the evolving constants to self-adjoint operators.

\subsection{Classical dynamics.}

Let us start with the classical action of the model.

\begin{equation} 
S=\int d\tau [p.{\dot u}+ \pi.{\dot v}  -  NH_1-MH_2-\lambda D]
\end{equation}

Where

\begin{equation} 
H_1= {1\over 2}(p^2-v^2),\;\;
H_2= {1\over 2}(\pi^2-u^2),\;\;
D=u.p-v.\pi;
\end{equation}

the dynamical variables $u=(u^1,u^2)$ and $v=(v^1,v^2)$ are
two-dimensional real vectors; $N$, $M$ and $\lambda$ are
Lagrange multipliers, and $u^2=u.u= (u^1)^2+(u^2)^2$. The
momenta conjugate to $u$ and $v$ are $p$ and $\pi$ and the
dynamics of the system is completely given by the constraints
$H_1$, $H_2$ and $D$. This is a totally constrained system with
three first-class constraints and a single degree of freedom.
The Poisson algebra of the constraints is

\begin{equation}
\{H_1,H_2\}=D\;\;
\{H_1,D\}= -2H_1\;\;
\{H_2,D\}= 2H_2
\end{equation}

which is the algebra sl(2,R). We now consider the observables.
The six functions [\cite{MoRoTh}]

\begin{eqnarray}
O_{12}&:=&u^1p^2-u^2p^1,\quad O_{23}:=u^2v^1-p^2\pi^1, \nonumber \\ 
O_{13}&:=&u^1v^1-p^1\pi^1,\quad O_{24}:=u^2v^2-p^2\pi^2, \nonumber \\
O_{14}&:=&u^1v^2-p^1\pi^2,\quad O_{34}:=v^2\pi^1-v^1\pi^2, 
\end{eqnarray}

commute with the constraints and form a closed Poisson bracket
algebra $o(2,2)$. This algebra is isomorphic to the Lie algebra
$sl(2,R)\times sl(2,R)$.  The combinations
 
\begin{eqnarray}
{\tau^\eta}_0&:=&{1\over2}(O_{12}-\eta O_{34}),\nonumber\\
{\tau^\eta}_1&:=&{1\over2}(O_{13}-\eta O_{24}),\nonumber\\
{\tau^\eta}_2&:=&{1\over2}(O_{23}-\eta O_{14}).
\end{eqnarray}

with $\eta \in \{1,-1\}$, are the elements of the basis adapted
to the algebra $sl(2,R)\times sl(2,R)$ with Poisson brackets

\begin{eqnarray}
\{{\tau^\eta}_1, {\tau^{\eta'}}_2\}&=&
-\delta^{\eta,\eta'}{\tau^\eta}_0, 
\nonumber\\
\{{\tau^\eta}_2, {\tau^{\eta'}}_0\}&=&
\delta^{\eta,\eta'}{\tau^\eta}_1, 
\nonumber\\
\{{\tau^\eta}_0, {\tau^{\eta'}}_1\}&=&
\delta^{\eta,\eta'}{\tau^\eta}_2,
\end{eqnarray}

Since the physical phase space is two-dimensional and the system
has one degree of freedom, there are at most two independent
continuous observables. In fact, one can show that \cite{MoRoTh}
the reduced physical phase space has the topology of four cones
connected at their vertices and all the observables evaluated at
regular points of each of the cones can be parametrized
\cite{LoRo} by

\begin{eqnarray}
{\tau^\eta}_0&=& {1\over2}\epsilon_1(1+\eta\epsilon_2)r ,\nonumber\\
{\tau^\eta}_1&=& {1\over2}(1+\eta\epsilon_2)r \cos{\phi} ,\nonumber\\
{\tau^\eta}_2&=& -{1\over2}\epsilon_1(1+\eta\epsilon_2)r \sin{\phi}.
\label{param}
\end{eqnarray}

where $r$ is a positive real parameter, $\phi$ is an angle, and
$\epsilon_1,\epsilon_2=\pm 1$ are two discrete quantities. The
Poisson brackets between $r$ and $\phi$ in the reduced phase
space reads

\begin{equation}
\{r,\phi\}=1
\end{equation}

while $\epsilon_1$ and $\epsilon_2$ commute with everything. The
functions on the phase space ${\tau^\eta}_i$ satisfy for each
$\eta$ the identity

\begin{equation}
-({\tau^\eta}_0)^2+({\tau^\eta}_1)^2+({\tau^\eta}_2)^2=H_1H_2+{1\over4}D^2 
\label{clasica}
\end{equation}

which vanishes on the physical phase space. The evolving
constants may be easily determined from the following identity
among the six observables $O_{ij}$ and the Lagrangian variables
${\vec u},{\vec v}$

\begin{equation}
u^a(\tau) v^b(\tau) \epsilon_{ac}\epsilon_{bd}(u^c(\tau)v^d(\tau)-
p^c(\tau)\pi^d(\tau))=O_{12}O_{34}\label{identi}
\end{equation}

where $\epsilon_{ab}$ are the Levi Civita symbols. By noticing
that each component of $u^cv^d-p^c\pi^d$ corresponds to one
observable, and making use of
 parameterisation  (\ref{param}) this relation takes the form

\begin{equation}
(\epsilon_2u^1(\tau)v^1(\tau)-u^2(\tau)v^2(\tau))\cos{\phi}-
(u^1(\tau)v^2(\tau)+
\epsilon_2u^2(\tau)v^1(\tau))\epsilon_1\sin{\phi}=r
\label{ident1} 
\end{equation}

The Lagrangian variables obey this relation at any time. From
here one can define an evolving constant $U^1$ that takes the
value $u^1$ when $u^2$ and $\vec v$ have assigned values:
$u^2=x,v^1=y,v^2=z$. It is given by

\begin{equation}
U^1={{x(z\cos{\phi}+\epsilon_2\epsilon_1y\sin{\phi})+r}\over{\epsilon_2y
\cos{\phi} - z\epsilon_1\sin{\phi}}}
\label{evolv}
\end{equation}
 
By considering the gauge orbits of Eq(\ref{ident1}) generated by
the constraints $H_1,H_2$ and $D$ one can obtain \cite{Mo} the
remaining evolving constants.

\begin{eqnarray}
P_1&=&\epsilon_2y\sin{\phi}+\epsilon_1z\cos{\phi}\\
P_2&=&-z\sin{\phi}+\epsilon_1\epsilon_2y\cos{\phi}\\
\Pi_1&=&{{xy+\epsilon_1\epsilon2r\sin{\phi}}\over{\epsilon1
\epsilon_2y\cos{\phi} 
- z\sin{\phi}}}\\
\Pi_2&=&{{xz+r\cos{\phi}}\over{\epsilon1\epsilon_2y\cos{\phi} - 
z\sin{\phi}}}\label{evolv1}
\end{eqnarray}

Thus, the relational description of this system involves three
independent variables $x,y,z$. Notice that not all the classical
evolving constants are independent. In fact, given $U^1$ and its
conjugate momentum $P_1$, the other quantities are determined up
to a sign factor by the constraints. In what follows, we shall
discuss if, at least for some choice of clock variables, the
evolving constants may be promoted to self-adjoint operators on
some Hilbert space leading to a consistent description of the
quantum evolution.

\subsection{Algebraic quantization.}

The SL(2,R) model has been previously quantized by following the
algebraic quantization approach. However, it has not been
possible up to now to give a satisfactory description of the
quantum evolution. We shall briefly describe this procedure
following closely the analysis of \cite{MoRoTh,LoRo} and we
shall explore whether the evolving constants may be included in
the algebra of observables ${\cal B}^{{(*)}}_{phys}$.

One works in a "coordinate representation" with wavefunctions
$\psi({\vec u},{\vec v})$, and considers the *-algebra of
physical observables ${\cal B}^{(*)}_{phys}$ generated by the
real operators ${{\hat \tau}^\eta}_i$. If one defines ${{\hat
\tau}^\eta}_\pm:= {{\hat \tau}^\eta}_1 \pm i{{\hat
\tau}^\eta}_2$ they satisfy the following commutation relations

\begin{equation}
[{{\hat \tau}^\eta}_0,{{\hat \tau}^{\eta'}}_\pm]=
\pm\delta^{\eta,\eta'}{{\hat 
\tau}^\eta}_\pm, \label{comm1}
\end{equation}

\begin{equation} 
[{{\hat \tau}^\eta}_{+},{{\hat \tau}^{\eta'}}_{-}]=
-2\delta^{\eta,\eta'}{{\hat 
\tau}^\eta}_0, \label{comm2}
\end{equation}
 
One can solve the quantum constraints by separation of variables
\cite{MoRoTh}. For this purpose, it is convenient to introduce
polar coordinates defined by
$u^1+iu^2=ue^{i\alpha},\;v^1+v^2=ve^{i\beta}$ One finds the set
of smooth solutions given by:

\begin{equation}
\psi_{m,\epsilon}:= e^{im(\alpha +i \epsilon \beta)}J_m(uv)
\end{equation}

where $m$ is an integer, $\epsilon \in \{-1,1\}$, and $J_m$ is
the Bessel function of the first kind. Notice that the functions
$\psi_{m,\epsilon}$ are linearly independent with the exception
that $\psi_{0,+}=\psi_{0,-}$. One can represent the action of
the generators of the *-algebra on the linear space spanned by
$\psi_{m,\epsilon}$ by

\begin{eqnarray}
{{\hat \tau}^\eta}_0 \psi_{m,\epsilon}=
\delta^{\eta,\epsilon}m\psi_{m,\epsilon}
\nonumber\\
{{\hat \tau}^\eta}_\pm \psi_{m,\epsilon}=
\delta^{\eta,\epsilon}m\psi_{m \pm 
1,\epsilon}\label{realiza}
\end{eqnarray}

The *-relations induce the adjoint operations

\begin{eqnarray}
({{{\hat \tau}^\eta}_0})^\dagger &=& 
{{{\hat \tau}^\eta}_0} \nonumber\\
({{{\hat \tau}^\eta}_\pm})^\dagger &=& 
{{{\hat \tau}^\eta}_\mp}\label{daga}
\end{eqnarray}

Now one can define an inner product among the elements of this
linear space.  The eigenfunctions $\psi_{m,\epsilon}$ of the
self-adjoint operator ${{\hat \tau}^\eta}_0$ should be
orthogonal, and taking into account equations (\ref{realiza})
and (\ref{daga}) one gets a recurrent relation between
$(\psi_m,\psi_m)$ and $(\psi_{m\pm 1},\psi_{m \pm 1})$ that
allows one to determine the inner product up to a positive
constant $a$

\begin{equation}
(\psi_m,\psi_{m'})= a|m|\delta_{m,m'}.
\end{equation}

Notice that the states $\psi_{0,\epsilon}$ have vanishing norm.
As this state is annhilated by every operator in ${\cal
B}^{{(*)}}_{phys}$, it can be excluded from the physical space.
One can thus define four linear spaces
$V_{\epsilon_1,\epsilon_2}:=\hbox{\rm
span}\{\psi_{m,\epsilon_2}: \epsilon_1 m >0\}$. Each of them
carries an irreducible representation of ${\cal
B}^{(*)}_{phys}$. The completion of these spaces yields the four
Hilbert spaces ${\cal H}_{\epsilon_1,\epsilon_2}$.  Finally,
notice that the Casimir operators $({\tau^\eta})^2= - ({{\hat
\tau}^\eta}_0)^2 + ({{\hat \tau}^\eta}_1)^2 +({{\hat
\tau}^\eta}_2)^2$ vanish on each Hilbert space ${\cal
H}_{\epsilon_1,\epsilon_2}$ and therefore, the quantum theory
preserves the classical identity (\ref{clasica}).

Now, if one attempts to analyse the problem of the relational
description of the evolution in the context of the algebraic
quantization on ${\cal H}_{\epsilon_1,\epsilon_2}$ one find
obstacles from the outset. For that analysis, we would like to
include among the elements of the algebra of perennials at least
some of the evolving constants. One should include in ${\cal
B}^{(*)}_{phys}$ the independent evolving constants $U^1$ and
$P_1$. Recall that at the classical level they allow us to
determine the other evolving constants on the constraint
surface. One can easily check that $P_1$ and $U^1$ are not well
defined self-adjoint operators on ${\cal
H}_{\epsilon_1,\epsilon_2}$. In fact now the troublesome vectors
$\psi_{0,\epsilon}$ of vanishing norm are not annihilated by the
new elements of the algebra and therefore they cannot be dropped
at the outset. Furthermore, if one attempts to restrict the
action of these operators to one of these sectors, the inverse
of $P_2$ appearing in the definition of $U^1$ is not defined on
any sector of ${\cal H}_{\epsilon_1,\epsilon_2}$. Therefore, the
quantum theory resulting from the algebraic quantization does
not seem to admit a natural time structure induced by the
evolving constant formalism.

\subsection{Reduced phase space quantization and relational
evolution of the SL(2,R) model.}

We shall see in this subsection that contrary to what we have
observed previously, it is possible to find in this case a
consistent description of the time evolution. It will be
convenient to begin by introducing a reparameterisation of the
reduced phase space of the model. We start from Eq(\ref{param})
and define $J=\epsilon_1 r$ and $\varphi=-\epsilon_1\phi+
{(1+\epsilon_1)\over 2} \pi$.  This parameterisation allows us
to rewrite the generators of the $sl(2,R)$ algebra as follows:

\begin{eqnarray}
{\tau^\eta}_0&=& {1\over2}(1+\eta\epsilon_2)J ,\nonumber\\
{\tau^\eta}_1&=& -{1\over2}(1+\eta\epsilon_2)J \cos{\varphi} ,
\nonumber\\
{\tau^\eta}_2&=& -{1\over2}(1+\eta\epsilon_2)J \sin{\varphi}.
\label{reparam}
\end{eqnarray}

The Poisson brackets between $J$ and $\varphi$ in the reduced
phase space read

\begin{equation}
\{\varphi, J\}=1 \label{varPoi}
\end{equation}

This new parametrization allows us to completely absorb the
$\epsilon_1$ factors and therefore, for each value of
$\epsilon_2$, describes simultaneously two opposite cones and
the common tip of the reduced phase space\cite{MoRoTh}. One can
easily check that the complete set of evolving constants
(\ref{evolv}) and (\ref{evolv1}) may be written without any
reference to $\epsilon_1$. Thus, within this parameterisation,
the natural subsets of the reduced phase space are not smooth
manifolds. This will have important consequences at the quantum
level.  We are interested in finding a quantum representation,
including in the fundamental *-algebra of observables -besides
the generators of the $sl(2,R)$ algebra- the evolving constants
$U_1(x,y,z)$ and $P_1(y,z)$, at least for some choice of clock
variable.  When written in terms of the new parameters, the
evolving constants take the form:

\begin{equation}
U^1={{x(z\cos{\varphi}-\epsilon_2y\sin{\varphi})-
J}\over{\epsilon_2y\cos{\varphi} + z\sin{\varphi}}}
\label{evolvi}
\end{equation}
 
\begin{equation}
P_1=\epsilon_2y\sin{\varphi}-z\cos{\varphi}
\end{equation}

These quantities take a more compact form by considering the
following change of the independent coordinates
$y=\epsilon_2\rho \cos{\varphi_0}$ and $z=-\rho \sin{\varphi_0}$

\begin{eqnarray}
P_1&=&\rho\sin(\varphi+\varphi_0)\\
U^1&=&-{x\rho\sin(\varphi+\varphi_0)+J}\over{\rho 
\cos(\varphi+\varphi_0)}
\end{eqnarray}

Now we are ready to quantize this system. We consider a linear
space of functions $\psi(\varphi)$. We then promote the Poisson
bracket relation (\ref{varPoi}) to a commutator by taking ${\hat
\varphi}$ as a multiplicative operator and ${\hat J}=-i{\partial
\over \partial \varphi}$ and define the quantum generators of
the sl(2,R) algebra by making use of the parametrization
(\ref{param}) as:

\begin{eqnarray}
{{\hat \tau}^\eta}_0&=& {1 \over 2}(1+\eta\epsilon_2){\hat J} 
\nonumber\\
{{\hat \tau}^\eta}_\pm&=& {-1\over 4}(1+\eta\epsilon_2)
[{\hat J} e^{\pm 
i\varphi} + e^{\pm i\varphi}{\hat J}]\label{taus} 
\end{eqnarray}

One can immediately check that they satisfy the commutation
relations (\ref{comm1}) and (\ref{comm2}) and therefore the
$sl(2,R)$ algebra. Since we are working directly on the reduced
phase space of perennials we may wonder if the quantum theory
preserves the classical identity (\ref{clasica}). To answer this
question we compute the Casimir operators. One gets

\begin{equation}
({\tau^\eta})^2= - ({{\hat
\tau}^\eta}_0)^2 + 
{1\over 2}({{\hat \tau}^\eta}_+{{\hat \tau}^\eta}_- +
{{\hat \tau}^\eta}_-{{\hat \tau}^\eta}_+) = 
{1\over 8}(1+\eta\epsilon_2)
{\hbar}^2, 
\end{equation}

Thus, we see that the quantization of the reduced phase space,
given by the singular manifold composed by two opposite cones
with a common tip, introduces a quantum anomaly and the Casimir
invariant only vanishes in the limit ${\hbar} \rightarrow 0$.

As we have already mentioned, we are interested in including
among the algebra of observables the evolving constants ${\hat
P}_1$ and ${\hat U}^1$ on the linear space of functions of
$\varphi$. The first operator has a trivial, multiplicative
action on this space, while the natural candidate for the second
is given by the symmetric combination:

\begin{equation}
{\hat U}^1\psi(\varphi) =-\{{1\over 2}[{\hat J}{1\over{\rho 
\cos(\varphi+\varphi_0)}} + {1\over{\rho
\cos(\varphi+\varphi_0)}} {\hat J}]
+{x \tan(\varphi+\varphi_0)}\}\psi(\varphi)
\end{equation}

In order to introduce a time structure in the system, one needs
to analize several questions. Firstly, one needs to find an
inner product such that the *- algebra be preserved. In
particular, one needs to ensure that the real evolving constants
turn out to be, at the quantum level, self-adjoint operators on
the corresponding Hilbert space. Secondly, one needs to analize
the issue of the clock variables in the case of multidimensional
orbits. Is it possible to consider all the independent
coordinates $x, \rho, \varphi_0$ as time variables, or do some
of them play a different role? If there is a time structure, is
that structure again related with the existence of surfaces
transversal to the orbits or may this condition be somewhat
relaxed in this case?

Before being more specific with the inner product, let us give a
first glance to the issue of the clock variables.  We first
notice that, being $P_1$ a multiplicative operator, its
corresponding eigenfunctions will have the form
$\psi_{\varphi_1}(\varphi)=\delta(\varphi- \varphi_1)$, and
therefore they do not depend on the independent coordinates that
label the evolution along the orbits. In what concerns the
evolving constant $U^1$, its eigenvalue equation is

\begin{equation}
{\hat U}^1\psi_{u^1}(\varphi) = u^1 \psi_{u^1}(\varphi)\label{eigenu}
\end{equation}

with solutions

\begin{equation}
\psi_{u^1}(\varphi) = A \sqrt{|\rho\cos(\varphi+\varphi_0)|}
\exp -i[u^1 
\rho\sin(\varphi+\varphi_0)+x\rho\cos(\varphi+\varphi_0)]
\label{eigenfunc}
\end{equation}

Since we have not introduced an inner product, we do not know
which of these solutions appear in the spectral decomposition of
${\hat U}^1$. However, notice that $x$ seems to play the role of
a time variable, in the sense that the transition amplitudes
$<\psi_v|\psi_u>$, computed with any sesquilinear inner product,
will be "time" independent. If that is the case, then
$\cos(\varphi+\varphi_0)$ will play the role of the "energy" of
the system.

One can easily check that the standard inner product of square
integrable functions of $\varphi$ in the interval $[-\pi,\pi]$
does not lead to a self-adjoint evolving constant ${\hat U}^1$.
However, if one divides the interval in two, and considers the
Hilbert space ${\cal H}_{\varphi_0}$ of periodic functions in
the interval $-\pi/2 \leq \varphi+\varphi_0 < \pi/2$ with inner
product

\begin{equation}
<\psi|\psi'>= 
{1\over \pi} \int^{{+\pi\over2}-\varphi_0}_{{-\pi\over2}-
\varphi_0} d\varphi \psi^*(\varphi)\psi'(\varphi)\label{innerred}
\end{equation}

which corresponds to the sector of positive "energy", the
elements of the quantum algebra ${\cal B}^{(*)}_{phys}$ are
self-adjoint on ${\cal H}_{\varphi_0}$. In order to prove that
the generators of the $sl(2,R)$ algebra ${{\hat \tau}^\eta}_i$
are self-adjoint, it is enough to take into account
Eq.(\ref{reparam}) and to show that $\hat J$ is self-adjoint.
This results from the fact that $\hat J$ admits the following
spectral decomposition in ${\cal H}_{\varphi_0}$

\begin{equation}
<\varphi|\hat J|\varphi'>= 
\sum_p {2p\over\pi}e^{i2p(\varphi-\varphi')} = -
i{\partial\over \partial \varphi} \delta_\pi(\varphi-\varphi') 
\end{equation}

where $\delta_\pi$ is the periodic Dirac delta. In what concerns
the evolving constants, ${\hat P}_1(\rho,\varphi_0)$ is
trivially self-adjoint while ${\hat U}^1(x,\rho,\varphi_0)$ has
the following set of orthonormal eigenvectors:

\begin{equation}
<\varphi|n,x\rho>= \sqrt{\pi/2} \sqrt{|\cos(\varphi+\varphi_0)|}
\exp i[n\pi \sin(\varphi+\varphi_0)+x\rho 
\cos(\varphi+\varphi_0)],\label{eigenu1}
\end{equation}

corresponding to the eigenvalues ${u^1}_n ={n \pi \over \rho}$.
The solutions of Eq.(\ref{eigenu}) for the other values of $u^1$
do not belong to the domain of the symmetric operator $U^1$.
\footnote{Recall that $A$ is said symmetric if, its domain $D_A$
is dense in ${\cal H}$ and for $f,\; g \in D_A$, (Af,g)=(f,Ag).}
Notice that, while the operator ${\hat U}^1$ has a continuum
spectrum on the kinematical space of functions $\psi(u,v)$ its
counterpart has a discrete spectrum on the reduced physical
space. This eigenfunctions define a complete basis, in fact

\begin{equation}
\delta_\pi(\varphi-\varphi') = 
{1\over\pi}\sum_n{<\varphi|n,x\rho><n,x\rho|\varphi'>},
\end{equation}

and therefore, $U^1$ is self-adjoint in ${\cal H}_{\varphi_0}$
with spectral decomposition

\begin{equation}
<\varphi'|U^1|\varphi> = \sum_n {{n\pi^2}\over 
{2\rho}}\sqrt{\cos\varphi'\cos\varphi}
e^{in\pi(\sin\varphi-\sin\varphi')}
\end{equation}

Thus, the reduced phase space quantization leads to a
satisfactory description of the time evolution of the $SL(2,R)$
model. The variable $T=x\rho$ plays the role of a clock
variable. The inner product between eigenstates of the evolving
constant $U^1$ is time independent, that is $<n,T|n',T> =
<n,T'|n',T'>$ which corresponds to a unitary evolution.
Furthermore, it is immediate to check from the definition of the
inner product (\ref{innerred}), that the transition
probabilities $<n,T|n',T'>$ are independent of $\varphi_0$. In
other words, all the Hilbert spaces ${\cal H}_{\varphi_0}$ give
equivalent descriptions of the evolution. Also notice that the
transition amplitudes are independent of $\rho$, the original
dependence on these parameter of the eigenfunctions of ${\hat
U}^1$ given by Eq.(\ref{eigenfunc}) was absorbed in the
normalization and the only remnant dependence on this parameter
appears in the spectrum of the evolving constant.  This
ambiguity in the spectrum may be easily understood by noticing
that by considering the partial gauge fixing $v^1={v^1}_0, \,
v^2={v^2}_0$ the original model may be interpreted as a
constrained system with Hamiltonian constraint \footnote{Notice
that after a partial gauge fixing, the sl(2R) model takes the
form of a Barbour Bertotti \cite{BB} model. Our description of
the quantum evolution naturally includes these models.}
$(p_1)^2+(p_2)^2= \rho^2$. The parameter $\rho$ plays the role
of the mass of this constraint system. Different mass values
obviously correspond to different representations of the
*-algebra of observables.  Notice that the equal time surfaces
are not transversal to the orbits. By using the classical form
of the evolving constants in terms of $\varphi$ and $J$ one can
see that the orbit corresponding to the value of $\varphi= -{\pi
\over 2} -\varphi_0$ lays completely on the equal time surface
$T= J$.  Thus, it is possible to get a satisfactory description
of the "time" evolution of the SL(2R)  model in spite of the
fact that the equal time surfaces are not globally transversal
to the orbits. Notice, however, that although the equal-time
surfaces contain causally connected points, for the resulting
quantum theory the probability amplitudes $<\pm{\pi \over 2}
-\varphi_0|n,T>$ vanish for any value of $n$. If one excludes
these pathological orbits, all the remaining orbits are
transversal to the equal time surfaces.

The region of the reduced phase space described by a particular
choice of $\varphi_0$ corresponds to $J\in [-\infty,\infty];
\varphi+\varphi_0 \in [- \pi/2,\pi/2)$ and covers one half of
the two opposite cones, for each value of $\epsilon_2$. The
dynamics of the other half may be easily included.  It is enough
to consider wave functions with two components
$\psi_a(\varphi)\; a=1,2$ with $\varphi+\varphi_0 \in
[-\pi/2,\pi/2)$ and inner product

\begin{equation}
<\psi|\psi'>= 
{1\over \pi} \int^{{+\pi\over2}-\varphi_0}_{{-\pi\over2}-
\varphi_0} d\varphi 
\sum_a {{\psi^*}_a(\varphi){\psi'}_a(\varphi)}.\label{innerreda}
\end{equation}

The first component corresponds to the sector with positive
energy, while the second corresponds to the negative energy
sector.  The generators of the $sl(2R)$ algebra of observables
now take the form ${{\hat \tau}^\eta}_0 \otimes I$, ${{\hat
\tau}^\eta}_\pm \otimes I$ where the $\hat \tau$s are given by
Eq.(\ref{taus}) and $I$ is the identity in the 2-dimensional
space. Finally the evolving constants are also diagonal with
$P_{22}(\varphi,\rho,\varphi_0)=
P_{11}(\varphi+\pi,\rho,\varphi_0)$ and
${U^1}_{22}(J,\varphi,x,\rho,\varphi_0)=
{U^1}_{11}(J,\varphi+\pi,x,\rho,\varphi_0)$.

To conclude, we have shown that it is possible to introduce a
consistent notion of quantum evolution in a system with more
than one Hamiltonian constraint. Dirac's quantization and
reduced phase space quantization lead in this case to
unequivalent quantum theories. Only the last procedure allowed
us to introduce a time structure. However, one cannot discard
the possibility of finding a consistent notion of evolution
within the context of the Dirac's quantization.

\section{A model with a compact presymplectic phase space}

In this section we are going to consider a system with a compact
constraint surface in phase space. This system does not contain
any $\Sigma\times R$ structure, and therefore it does not
correspond to a Hamiltonian system. It was first analysed by
Rovelli\cite{Ro42}. He concluded that although in a certain
regime, which stands for certain states and a certain range of
measurements, the system behaves as a standard quantum
mechanical system, while outside this regime unitarity is lost.  
The lessons learnt in this paper will allow us to introduce a
time structure that will preserve the unitarity of the evolution
and the standard probabilistic interpretation.

The model is given by a phase space with canonical coordinates
$q_1,q_2,p_1,p_2$ and canonical commutation relations
$\{q_i,p_j\}=\delta_{ij}$.  The constraint is
 
\begin{equation}
C={1 \over 2}({p_1}^2+{p_2}^2+{q_1}^2+{q_2}^2)-M.
\end{equation}

The equations of motion may be easily integrated leading to the
parametric equations for the orbits

\begin{eqnarray}
q_1&=&\sqrt{{2A}}\sin{(\tau)}\nonumber\\
q_2&=&\sqrt{2M-2A}\sin{(\tau+\phi)}
\end{eqnarray}
 
The orbits are labelled by the integration constants 

\begin{equation}
0 \le A \le M \;\; a \le \phi \le 2\pi
\end{equation}

For $\pi \le \phi \le 2\pi$ one has the same orbits as for $0
\le \phi \le \pi$, but followed with the opposite orientation.  
Geometrically, the orbits are ellipses that are inscribed in a
rectangle with a diagonal equal to M. The observables $A$ and
$\phi$ form a complete set and the topology of the reduced phase
space has the topology of a sphere. The explicit form of the
observables in terms of the original phase space variables is

\begin{eqnarray}
A&=&{1\over4}(2M+{p_1}^2-{p_2}^2+{q_1}^2-{q_2}^2)\nonumber\\
\tan{\phi}&=& {{p_1q_2-p_2q_1}\over{p_1p_2+q_1q_2}}
\end{eqnarray}

The relational evolution of some of the coordinates in terms of
the others may be easily determined from the following identity
among the two obsevables and the Lagrangian variables ${\vec q}$

\begin{equation}
(\mu q_2- \cos{\phi}q_1)^2 +\sin^2{\phi}q_1^2= 2A \sin^2{\phi}
\end{equation}

where $\mu =\sqrt{A/M-A}$. If we simply try to describe the
evolution of one of the coordinates, say $q_2$ in terms of $q_1$
that will play the role of clock variable, one runs into
problems \cite{Ro42}. In fact,

\begin{equation}
q_2(q_1)=\sqrt{M/A-1}[\cos\phi q_1\pm \sin\phi(2A-{q_1}^2)^{1/2}]
\end{equation}

and we see that due to the square root, the family of orbits
going trough a certain equal time surface will depend on
$q_1=T$. As $q_1$ grows, less orbits reach the "equal-time"
surface. At the quantum level this implies that, the evolving
constant cannot be self-adjoint operators, and the evolution
will not be unitary.

We shall see that there are, however, choices of clock variables
and evolving constants, leading to self-adjoint observables and
unitary evolutions. The main observation is that, as for the
model analysed in the previous section, it is possible to define
clock variables such that the equal time surfaces, up to
isolated pathological orbits, are transversal to the orbits. We
shall call this kind of surfaces locally transversal. To do
that, we define new configuration variables $r$ and $\alpha$
such that $q_1=r\sin\alpha$ and $q_2=r\cos\alpha$, and choose
the clock variable $T=\alpha$. The evolving observable

\begin{equation}
R(T) =
{\sqrt{2A\sin^2{\phi}}\over[(\mu\cos\alpha-\cos\phi\sin\alpha)^2+
\sin^2\phi\sin^2\alpha]^{1/2}} 
\end{equation}

defines a one parameter set of physical observables. Except for
a couple of orbits which lay completely on the constraint
surface, all the orbits are transversal to the equal time
surfaces. Looking at the zeros of the denominator one can detect
the pathological orbits; they are $\phi=0$ laying on the equal
time surfaces $\tan\alpha= \mu$ and $A=0$ laying on the surfaces
$\alpha=n\pi$. These orbits correspond to the case where the
ellipses degenerate to a straight line.  In order to quantize
the system, it will be convenient to introduce, following
Rovelli\cite{Ro42}, the observables

\begin{eqnarray}
L_x&=&{1\over2}(p_1p_2+q_1q_2),\nonumber\\
L_y&=&{1\over2}(p_1q_2-q_1p_2),\nonumber\\
L_z&=&{1\over2}({p_1}^1-{p_2}^2 + {q_1}^2 -{q_2}^2),
\end{eqnarray}

which are related by

\begin{equation}
{L_x}^2+{L_y}^2+{L_z}^2= {M^2 \over 4}
\end{equation}

In terms of the original variables $A$ and $\phi$, they take the
following form

\begin{eqnarray}
L_x&=& \sqrt{A(M-A)}\cos\phi,\nonumber\\
L_y&=& \sqrt{A(M-A)}\sin\phi,\nonumber\\
L_z&=& A-{M\over2}
\end{eqnarray}

and obey the angular momentum Poisson algebra.

We are interested in the quantization of the *-algebra of
observables generated by the observables $\vec L$ and the
evolving constant $R(T)$. The quantization of the observables
$\vec L$ is straightforward, using the standard representations
of the su(2)algebra. It follows that

\begin{equation}
{\hat L}^2= {M^2\over4}= j(j+1)
\end{equation}

where $j$ is integer or half-integer, and therefore the
classical limit is preserved. We shall see in what follows that
only the half-integer values of $j$ are admissible.  The
corresponding Hilbert space will have a standard basis $|m>$ of
eigenvectors of $\hat L_z$, with $-j\le m\le j$. In order to
have a well defined description of the evolution at the quantum
level, we need to promote $R(T)$ to a self-adjoint operator. As
we are working on a finite dimensional Hilbert space, it is
enough to show that ${\hat R}(T)$ is a well defined symmetric
operator. Thus, we need to show that the pathological orbits
$A=0$ and $\sin\phi=0$ have vanishing probabilities.

Notice that

\begin{equation}
\hat A|m> (M/2+m)|m> = (\sqrt{j(j+1)}+m)|m> \not= 0  
\end{equation}

for all $m$ such that $-j\le m\le j$, and therefore $A=0$ is not
an eigenvalue of $\hat A$. Concerning the possible values of
$\sin\phi$, we notice that it takes the following form, in terms
the ${\hat L}$
 operators.

\begin{equation}
\hat{\sin\phi}= 
(M^2/4-{\hat L_z}^2)^{-1/4}{\hat L}_x(M^2/4-{\hat L_z}^2)^{-1/4}
\end{equation}

This is a self-adjoint operator which for half-integer $j$ has a
well defined inverse. In fact, in the half-integer
representations ${\hat L}_x$ does not have vanishing eigenvalues
and is invertible. Therefore, zero is not included among the
possible eigenvalues of $\hat{\sin \phi}$. Thus, this quantum
description assigns zero probability to the pathological orbits.
From here, it is not difficult to prove that the symmetric form
of the evolving constants

\begin{eqnarray}
{\hat R}(T)=[({\hat L}_z+M/2)\cos\alpha + 
{\hat L}_x\sin\alpha)^2+
{{\hat L}_y}^2\sin^2\alpha]^{-1/4}(2{\hat L}_z+M)^{1/4} 
\sqrt{{{\hat L}_y}^2}\\
\times(2{\hat L}_z+M)^{1/4}[({\hat L}_z+M/2)
\cos\alpha + {\hat L}_x\sin\alpha)^2+{{\hat L}_y}^2
\sin^2\alpha]^{-1/4}
\end{eqnarray}

is a positive definite self-adjoint operator. To see this, it is
enough to check that the negative powers are well defined
because they involve positive definite self-adjoint operators.
Thus, if we label the eigenvalues of $R(T)$ by $r_n(T)$ and the
eigenvectors by $|n,T>$ they form an orthonormal basis of the
$2j+1$ dimensional Hilbert space, such that $<n,T|m,T>=
\delta_{n,m}$ for any $T$. The eigenvectors corresponding to
different $T$ are related by the unitary transformation
$U(n,T;n'T')=<n,T|n'T'>$ and one recovers the standard
probabilistic interpretation of the quantum theory. The only
peculiarity of this model lays in the fact that the eigenvalues
of $R(T)$ are "time dependent". If one uses the unitary
transformation to go from the Heisenberg representation to the
Schroedinger representation, one gets a time dependent
Schroedinger operator.

\section{Conclusions}

In this paper, we have addressed the issue of the relational
description of the evolution of generally covariant systems. We
have shown that Rovelli's evolving constants are useful tools
for the analysis of the evolution in terms of relational time
variables. Consistency with quantum mechanics and its
probabilistic interpretation puts stringent constraints to the
admissible choices of clock variables. The description in terms
of evolving observables allows us to compare the descriptions
arising from different time choices simply by introducing in the
fundamental *-algebra the corresponding evolving observables.
Generically, each of these choices corresponds to non-commuting
observables and different experimental setups. Therefore, the
probability amplitudes assigned to the same event by two
different choices will be, in general, also different. This was
already noticed by Hartle\cite{Har}, who concluded that the
probabilities of a history may differ due to the time choice.

We were able to give consistent quantum mechanical descriptions
for the evolution of systems with a compact reduced phase space
and for systems with more than one Hamiltonian constraint.
Although the orbits of these kind of systems depend on several
parameters, it was possible to describe them with a single clock
variable. Some of these systems do not admit equal time surfaces
globally transversal to the orbits. In fact, we have considered
systems such that most of the orbits are transversal, but they
have some exceptional isolated orbits laying completely on an
equal-time surface. We have shown that such systems may be
consistently quantized. The resulting quantum theory assigns a
vanishing probability to these pathological orbits.

To what extent can we generalize these conclusions to more
realistic relational systems? The models we have considered are
certainly very simple but certain tentative conclusions can be
drawn. The fundamental limitation to the choices of clock
variables seems to be the causal independence of the points
laying on an equal time surface. One can imagine two kinds of
extensions of this analysis compatible with this restriction. In
first place one may have systems in which for any choice of the
clock variable $Q$, the number of orbits reaching the equal time
surface $Q=T$ varies with $T$. For certain orbits, the question
"where is the dynamical variable $O$ when the clock variable
takes the value $T$?" has no answer, because $Q$ never reaches
the value $T$ in that orbit. In other words, in this kind of
systems only a "time dependent" portion of the reduced phase
space is covered at each time $T$\footnote{As it was noticed by
Rovelli \cite{Ro44}, the choice of time variable $T=q_1$ in the
model considered in the previous section leads to a system of
this type.}. At the quantum level, these systems lead naturally
to evolving Hilbert spaces and the computation of transition
amplitudes between eigenvectors of the evolving observables at
different values of the time variable is problematic.  One can
also extend this analysis to systems that do not admit a global
clock variable. In this case, one is able to find equal time
surfaces that are locally transversal to the orbits, only for
certain range of the time variable that does not cover the
entire evolution of the system. In this case, one needs more
than one clock variable to cover the complete evolution of the
system. We are now studying models having these two kinds of
behavior.

\end{document}